\documentclass[
reprint,
superscriptaddress,
 amsmath,amssymb,
 aps,
]{revtex4-1}

\usepackage{comment}
\usepackage{graphicx}
\usepackage{dcolumn}
\usepackage{bm}
\usepackage{enumitem}
\usepackage{threeparttable}
\usepackage{color}
\usepackage{multirow}

\newcommand{\fig}[1]{Fig.~\ref{#1}}
\newcommand{\tab}[1]{Table~\ref{#1}}
\newcommand{\Sec}[1]{Sec.~\ref{#1}}

\newcommand{\g}{$\gamma$}

\newcommand{\ran}{($\alpha$,n) }
\newcommand{\rann}{($\alpha$,2n) }

\newcommand{\HeHe}{$^3$He($\alpha$,$\gamma$)$^7$Be }
\newcommand{\Lip}{$^6$Li(p,$\gamma$)$^7$Be }

\begin{document}

\title{Experimental determination of the \HeHe reaction cross section above the $^7$Be proton separation threshold}

\author{\'A.~T\'oth}%
\affiliation{Institute for Nuclear Research (Atomki), Debrecen, Hungary}
\affiliation{University of Debrecen, Doctoral School of Physics, Debrecen, Hungary}

\author{T.~Sz\"ucs}%
\email{tszucs@atomki.hu}%
\affiliation{Institute for Nuclear Research (Atomki), Debrecen, Hungary}

\author{T.~N.~Szegedi}%
\affiliation{Institute for Nuclear Research (Atomki), Debrecen, Hungary}

\author{Gy.~Gy\"urky}%
\affiliation{Institute for Nuclear Research (Atomki), Debrecen, Hungary}

\author{Z.~Hal\'asz}%
\affiliation{Institute for Nuclear Research (Atomki), Debrecen, Hungary}

\author{G.\,G.~Kiss}%
\affiliation{Institute for Nuclear Research (Atomki), Debrecen, Hungary}

\author{Zs.~F\"ul\"op}%
\affiliation{Institute for Nuclear Research (Atomki), Debrecen, Hungary}

\begin{abstract}
\begin{description}
\item[Background] The \HeHe reaction plays a major role both in the Big Bang Nucleosynthesis producing the majority of the primordial $^7$Li, and in the pp-chain of solar hydrogen burning, where it is the branching point between the pp-I and pp-II,-III chains. As a few-nucleon system, this reaction is often used to validate ab-initio theoretical calculations and/or test R-matrix theory and code implementations. For the latter, experimental data in an extended energy range is of crucial importance to test the fit and extrapolation capabilities of the different codes.
\item[Purpose] The \HeHe reaction cross section has been measured by several groups up to the first resonance ($E_{c.m.} \approx 3$~MeV) in the reaction. However, only one dataset exists above the $^7$Be proton separation threshold measured in a narrow energy range ($E_{c.m.} = 4.0-4.4$~MeV). In this work we extend the available experimental capture cross section database to the energy range of known $^7$Be levels, where only particle scattering experiments are available for testing the models.
\item[Method] The activation method was used for the \HeHe reaction cross section determination. The experiment was performed using a thin-window gas cell with two high-purity Al foils as entrance and exit windows. The activity of the $^7$Be nuclei implanted in the exit/catcher foil was measured by detecting the yield of the emitted $\gamma$~rays using shielded high-purity germanium detectors.
\item[Results] New experimental \HeHe reaction cross section data were obtained for the first time in the $E_{c.m.}=4.3-8.3$~MeV energy region, corresponding to $E_x=5.8-10$~MeV excitation energies of $^7$Be. The new dataset with about 0.2 MeV step covers the energy range of known levels and particle separation thresholds. No prominent structures are observer around the $^7$Be levels. 
\item[Conclusions] The measured reaction cross section is slowly increasing with increasing energy in the range of  $E_x=6-8$ MeV from 10 $\mu$b to 13 $\mu$b. Above the $^6$Li$+p_1$ threshold, a decrease starts in the cross section trend and reaches a value of about 8 $\mu$b around $E_x=10$ MeV. The overall structure of the cross section suggest a broad resonance peaking around $E_x=7.5$ MeV $^7$Be excitation energy, with a width of 8 MeV.
\end{description}
\end{abstract}

\maketitle

\section{\label{sec:intro} Introduction}
The \HeHe reaction is of crucial importance in three different nuclear astrophysics scenarios. It is one of the branching reactions of the proton-proton (pp) chains in solar and stellar hydrogen-burning. More specifically, it is the initial reaction of the pp-II and pp-III chains. These chains are the source of a significant portion of the high-energy neutrinos emitted by the Sun \cite{Haxton13-ARAA}. Accurate estimates of the neutrinos produced in the Sun can be used to refine solar models \cite{Magg22-AA}, however the rate uncertainty of the branching reaction directly affects the modeled neutrino flux uncertainty.
The \HeHe is also an important reaction of element formation in the Big Bang Nucleosynthesis (BBN). The formation of $^7$Li happens mainly via the \HeHe reaction and the subsequent beta-decay of $^7$Be \cite{Fields11-ARNPS}.
Additionally, classical novae also play an important role in the galactic production of $^7$Li. The synthesis of the $^7$Be (which is transformed into $^7$Li nuclei) in the \HeHe reaction can also be observed in carbon-oxygen type novae.
$^7$Be is detected by spectroscopic methods and several simulations have been carried out on the amount of $^7$Be \cite{Starrfield20-AJ} (and further references therein).

In the relevant reaction energy range in stars, in the so-called Gamow window, direct experimental data are difficult to obtain because of the extremely low reaction cross sections in the attobarn range. Here experimental information can be gained via indirect methods, e.\,g. the Asymptotic Normalization Coefficient (ANC) of the reaction was determined by using transfer reaction \cite{Kiss20-PLB}, or the reaction cross section in the solar Gamow-window ($0.018-0.029$~MeV) was determined utilizing the measured solar neutrino fluxes and the predictions of the Standard Solar Model \cite{Takacs15-PRD}. The relevant energy range in classical novae ($0.05-0.2$~MeV) and the BBN Gamow-window ($0.1-0.5$~MeV) are somewhat higher in energy, up to which the reaction cross section is increasing exponentially, reaching the nanobarn range. With an enormous effort, the LUNA collaboration \cite{luna} was able to provide direct experimental data in this energy range with high precision \cite{Bemmerer06-PRL, Confortola07-PRC, Gyurky07-PRC, Costantini08-NPA}. The collaboration explored the energy range of $E_{c.m.} = 0.09-0.17$ MeV, with their deep underground settings, where the environmental background signals in the detectors are orders of magnitude lower than that can be achieved in an overground setup \cite{Caciolli09-EPJA}.

Many other modern datasets are available in the $E_{c.m.} = 0.3-3.1$ MeV energy range \cite{Singh04-PRL,Brown07-PRC,DiLeva09-PRL, Carmona-Gallardo12-PRC, Bordeanu13-NPA, Kontos13-PRC} proving the positive slope of the astrophysical \mbox{S-factor} towards higher energies.

In addition the \HeHe reaction cross section was measured around the proton separation energy of the compound $^7$Be nucleus in a narrow energy range of  $E_{c.m.} = 4.0-4.4$~MeV \cite{Szucs19-PRC}. In this energy range a positive parity level of $^7$Be was suggested from the \Lip reaction \cite{He13-PLB} but not confirmed later in any direct experiment \cite{Piatti20-PRC} or indirect work \cite{Kiss21-PRC}.

Because the energy range of the solar and stellar \HeHe reaction is not reachable by present day direct experimental technique, extrapolation to those energies are inevitable. For this purpose, one of the often used methods is the R-matrix analysis \cite{Azuma10-PRC}. With the rapid growth of computational power, recently multi-level, multichannel R-matrix codes became available \cite{Thompson19-EPJA}, using known level properties from different experiments to extrapolate the S-factor into unknown energy ranges. Most of the previous \HeHe R-matrix studies for this purpose used the low energy radiative capture datasets only, below the $^7$Be proton separation threshold \cite{deBoer14-PRC, Odell22-FP}. However, the extrapolations may benefit from new experimental datasets in previously unexplored energy region, where the reaction of interest as well as other reaction channels can be used to constrain their parameters.

To describe the low energy trend of the S-factor, broad positive parity states have to be assumed in the R-matrix fit \cite{deBoer14-PRC}. Such an assumption is reasonable, based on the fact that in its mirror nucleus $^7$Li, a broad structure was found in $\gamma$-scattering experiments \cite{Skopik79-PRC, Junghans79-ZPA, Munch20-PRC} around 7 MeV excitation energy. In the present work the corresponding energy range is addressed in $^7$Be.
In addition, recently the $^3$He+$^4$He scattering datasets were used to cross-validate several R-matrix codes \cite{Thompson19-EPJA}. An extension of that work could be the inclusion of radiative capture channels, which requires experimental data up to 20 MeV in $^7$Be excitation energy.
Answering also to this call, we provide here an experimental capture cross section dataset up to $E_x=10$ MeV.

The cross section of the \HeHe reaction can be measured with several methods. In the case of prompt \g-ray detection, the direct capture \g~rays and/or the secondary \g-s are detected. This method was used so far only below the first resonance. The main reason is that the angular distribution of the prompt \g~rays is affecting the deduced cross sections, this needs to be known with high precision, or shall be estimated to be a small correction. Since the angular distribution by now is known only from theoretical works, the latter requirement is fulfilled only far away from resonances. A recent attempt is made to experimentally determine the prompt \g-ray angular distribution \cite{Turkat19-SNC}, and further work is in progress, which may unlock the potential of these kinds of measurements for precise cross section determination.

An alternative method for the capture cross section determination is the direct detection of the $^7$Be recoils. Because of the technical challenges, this method was successfully applied only by one group so far utilizing the ERNA (European Recoil Separator for Nuclear Astrophysics) apparatus, resulting in a dataset up to, and covering, the first resonance \cite{DiLeva09-PRL}. Experiments were carried out using the DRAGON (Detector of Recoils And Gammas Of Nuclear reactions) recoil separator \cite{Sjue13-NIMA}, however, the results are still reported only as conference contributions \cite{Singh12-JPCS, CarmonaGallardo14-EPJWOC}.

The third method is the so-called activation \cite{Gyurky19-EPJA}. The $^7$Be reaction product is radioactive with a half-life of 53.22 days and 10.44\% of its decays lead to the first excited state of $^7$Li, which subsequently emits a \g~photon with an energy of 477.6 keV \cite{Tilley02-NPA}. By detecting this latter \g~ray, the number of reaction products, thus the reaction cross section can be deduced.
The activation method is free from the angular distribution effects, influencing the other two methods, thus can be safely used also in energy ranges, where only limited information is available about the levels.
Therefore in this work, the activation method is applied to determine the \HeHe reaction cross section in an energy range never investigated in this radiative capture reaction before. The new dataset spans the energy range of known $^7$Be levels and particle emission thresholds.

This paper is organized as follows: in \Sec{sec:exp} the experimental details are given, highlighting all the constituents of the cross section determination. In \Sec{sec:results} the data analysis and the experimental results are presented. Finally a summary is given in \Sec{sec:sum}.

\section{\label{sec:exp} Experimental details}

\subsection{\label{sec:cell} The gas-cell target}

In the present work, a thin-window gas-cell target was used, an updated version of those used in Refs. \cite{Bordeanu12-NIMA, Szucs19-PRC}. With this solution, differential pumping and calorimetric beam current measurement, which is often necessary for a windowless gas target \cite{Ferraro18-EPJA} can be avoided. The disadvantage of a window is the beam energy loss in the entrance foil, which is not problematic in our case due to the relatively high beam energies used. The nominal 10~$\mu$m thick aluminium foils used as entrance window cause $0.5-1$ MeV energy loss in the beam energy range (E$_\alpha$ = $11-20$~MeV) of our investigations. There were a few data-points measured with thinner ($\sim$7~$\mu$m) entrance foils to explore the excitation function with finer energy steps in the vicinity of known $^7$Be levels. 

The exact entrance foil thicknesses were determined by measuring the energy loss of passing $\alpha$ particles. The energy of $\alpha$-s emerging from a triple-isotope $\alpha$ source penetrating through the foil was measured. The thickness of the aluminium foil was then determined from the energy loss as described in the previous work \cite{Szucs19-PRC}. The statistical uncertainty of the thickness measurement was 0.3-0.5\%, while the stopping power uncertainty was taken into account as follows. In the energy range of $3.15-5.80$~MeV covering the initial $\alpha$ energies as emerged from the source and the decelerated ions as detected, there are several stopping power measurements, providing a handful of datasets \cite{Andersen77-PRA, Diwan15-NIMB, Santry86-NIMB, Raisanen91-REDS, Nakata69-CJP, Desmarais84-AJP, Trzaska18-NIMB, Hsu05-NIMB, Majackij88-UFZ} with different accuracies to be compared by the SRIM tables \cite{srim} used in the present calculations. The energy dependence of most of the datasets are well described by the SRIM curve, however scaling factors within their quoted accuracy shall be applied. The weighted mean of these scale factors amounts to 1.009, while its uncertainty is 1.0\%. Therefore the thicknesses calculated using the SRIM tables were multiplied by 0.991 to account for this effect, while the uncertainty of the stopping power is taken as the spread of the scale factors, thus 1\% added quadratically to the statistical uncertainty of the measurement.
The parameters of the different irradiations, together with the window thicknesses as target areal densities are summarized in \tab{tab:run}.

The 4.19-cm long cell was filled with high purity isotopically enriched (99.999\%) $^3$He gas. The foils were placed on O-rings at the entrance and exit of the cell secured and pressed by tantalum rings. A 12 mm diameter surface of the foil was exposed to the gas, which securely kept the pressure against the beamline vacuum. Before the irradiations, the cell was filled with up to 100~mbar of $^3$He gas. From this initial pressure, temperature and the known cell length, the surface density of the target nuclei was determined (see \tab{tab:run}) applying the ideal gas law.

In the experiments, 4 different thicknesses of exit foils (10, 15, 20, 25 um)  were used serving as catcher, depending on the energy of $^7$Be produced in the reaction. The $^7$Be energy is calculated from the kinematics, and simulations with the SRIM program \cite{srim} were done to determine the thickness of the catcher foil required at the given energies to ensure that the $^7$Be nuclei are stopped in the foil and do not pass through it. The thinnest available foils were then used to reduce the number of target atoms for possible parasitic reactions.

\begin{table}[tb]
\caption{Parameters of the entrance foils, targets and irradiations used in the experiment.}
\label{tab:run}
\center
\begin{ruledtabular}
\begin{tabular}{c c c c c }									
\multirow{2}{*}{$E_{\alpha}$}	& Entrance foil															& Target														& \multirow{2}{*}{$t_{irrad}$}	& Average \\
								& thickness																& areal density														&								& beam current \\

(MeV)							& ($\mu$m)																& $\left( \frac{10^{19}\mathrm{at}}{\mathrm{cm}^2} \right)$				& (h) 							& ($\mu$A)  \\
\noalign{\smallskip}\colrule\noalign{\smallskip}									
11.0	& \begin{tabular}{r@{\,$\pm$\,}l}	10.26			&	0.12 \end{tabular} & \begin{tabular}{r@{\,$\pm$\,}l}	1.07	&	0.03 \end{tabular} &	17.6	& 0.95	\\	
11.5	& \begin{tabular}{r@{\,$\pm$\,}l}	10.18			&	0.11 \end{tabular} & \begin{tabular}{r@{\,$\pm$\,}l}	1.08	&	0.03 \end{tabular} &	19.1	& 0.85	\\		
12.0	& \begin{tabular}{r@{\,$\pm$\,}l}	10.31			&	0.11 \end{tabular} & \begin{tabular}{r@{\,$\pm$\,}l}	0.83	&	0.02 \end{tabular} &	15.5	& 1.00	\\
12.0	& \begin{tabular}{r@{\,$\pm$\,}l} 	\phantom{1}6.83&	0.07 \end{tabular} & \begin{tabular}{r@{\,$\pm$\,}l}	1.01	&	0.03 \end{tabular} &	17.3	& 0.59	\\
12.5	& \begin{tabular}{r@{\,$\pm$\,}l}	10.48			&	0.12 \end{tabular} & \begin{tabular}{r@{\,$\pm$\,}l}	1.01	&	0.03 \end{tabular} &	21.9	& 0.91	\\	
12.5	& \begin{tabular}{r@{\,$\pm$\,}l}	\phantom{1}6.83&	0.07 \end{tabular} & \begin{tabular}{r@{\,$\pm$\,}l}	1.02	&	0.03 \end{tabular} &	16.0	& 0.63	\\
13.0	& \begin{tabular}{r@{\,$\pm$\,}l}	10.15			&	0.11 \end{tabular} & \begin{tabular}{r@{\,$\pm$\,}l}	1.01	&	0.03 \end{tabular} &	15.4	& 0.84	\\	
13.0	& \begin{tabular}{r@{\,$\pm$\,}l}	\phantom{1}7.15&	0.08 \end{tabular} & \begin{tabular}{r@{\,$\pm$\,}l}	1.03	&	0.03 \end{tabular} &	21.1	& 0.82	\\		
13.5	& \begin{tabular}{r@{\,$\pm$\,}l}	10.48			&	0.12 \end{tabular} & \begin{tabular}{r@{\,$\pm$\,}l}	1.05	&	0.03 \end{tabular} &	19.6	& 1.00	\\	
13.6	& \begin{tabular}{r@{\,$\pm$\,}l}	\phantom{1}7.15&	0.08 \end{tabular} & \begin{tabular}{r@{\,$\pm$\,}l}	1.02	&	0.03 \end{tabular} &	20.0	& 0.70	\\		
14.0	& \begin{tabular}{r@{\,$\pm$\,}l}	10.31			&	0.11 \end{tabular} & \begin{tabular}{r@{\,$\pm$\,}l}	0.81	&	0.02 \end{tabular} &	19.4	& 1.04	\\
14.5	& \begin{tabular}{r@{\,$\pm$\,}l}	10.15			&	0.11 \end{tabular} & \begin{tabular}{r@{\,$\pm$\,}l}	1.01	&	0.03 \end{tabular} &	17.5	& 0.86	\\	
15.0	& \begin{tabular}{r@{\,$\pm$\,}l}	10.37			&	0.12 \end{tabular} & \begin{tabular}{r@{\,$\pm$\,}l}	1.02	&	0.03 \end{tabular} &	26.0	& 0.80	\\
15.5	& \begin{tabular}{r@{\,$\pm$\,}l}	10.52			&	0.12 \end{tabular} & \begin{tabular}{r@{\,$\pm$\,}l}	1.01	&	0.03 \end{tabular} &	21.2	& 0.86	\\
16.0	& \begin{tabular}{r@{\,$\pm$\,}l}	10.31			&	0.11 \end{tabular} & \begin{tabular}{r@{\,$\pm$\,}l}	0.82	&	0.02 \end{tabular} &	 \phantom{1}9.9		& 1.20	\\
16.5	& \begin{tabular}{r@{\,$\pm$\,}l}	10.52			&	0.12 \end{tabular} & \begin{tabular}{r@{\,$\pm$\,}l}	1.01	&	0.03 \end{tabular} &	22.2	& 0.77	\\			
16.6	& \begin{tabular}{r@{\,$\pm$\,}l} 	\phantom{1}6.83&	0.07 \end{tabular} & \begin{tabular}{r@{\,$\pm$\,}l}	1.05	&	0.03 \end{tabular} &	22.0	& 0.58	\\
17.0	& \begin{tabular}{r@{\,$\pm$\,}l}	\phantom{1}7.15&	0.08 \end{tabular} & \begin{tabular}{r@{\,$\pm$\,}l}	1.12	&	0.03 \end{tabular} &	24.1	& 0.80	\\					
17.5	& \begin{tabular}{r@{\,$\pm$\,}l}	10.33			&	0.11 \end{tabular} & \begin{tabular}{r@{\,$\pm$\,}l}	1.02	&	0.03 \end{tabular} &	16.0	& 0.82	\\		
18.0	& \begin{tabular}{r@{\,$\pm$\,}l}	10.56			&	0.11 \end{tabular} & \begin{tabular}{r@{\,$\pm$\,}l}	1.08	&	0.03 \end{tabular} &	23.5	& 0.59	\\	
18.5	& \begin{tabular}{r@{\,$\pm$\,}l}	10.15			&	0.11 \end{tabular} & \begin{tabular}{r@{\,$\pm$\,}l}	1.02	&	0.03 \end{tabular} &	21.1	& 0.87	\\	
19.0	& \begin{tabular}{r@{\,$\pm$\,}l}	10.48			&	0.12 \end{tabular} & \begin{tabular}{r@{\,$\pm$\,}l}	1.01	&	0.03 \end{tabular} &	22.0	& 0.95	\\			
19.5	& \begin{tabular}{r@{\,$\pm$\,}l}	10.15			&	0.11 \end{tabular} & \begin{tabular}{r@{\,$\pm$\,}l}	1.02	&	0.03 \end{tabular} &	19.8	& 0.92	\\			
20.0	& \begin{tabular}{r@{\,$\pm$\,}l}	10.48			&	0.12 \end{tabular} & \begin{tabular}{r@{\,$\pm$\,}l}	1.02	&	0.03 \end{tabular} &	22.1	& 0.84	\\			
\end{tabular}									
\end{ruledtabular}						
\end{table}

\subsection{\label{sec:irrad} Irradiations}

\begin{figure}[t]
\includegraphics[width=0.95\columnwidth]{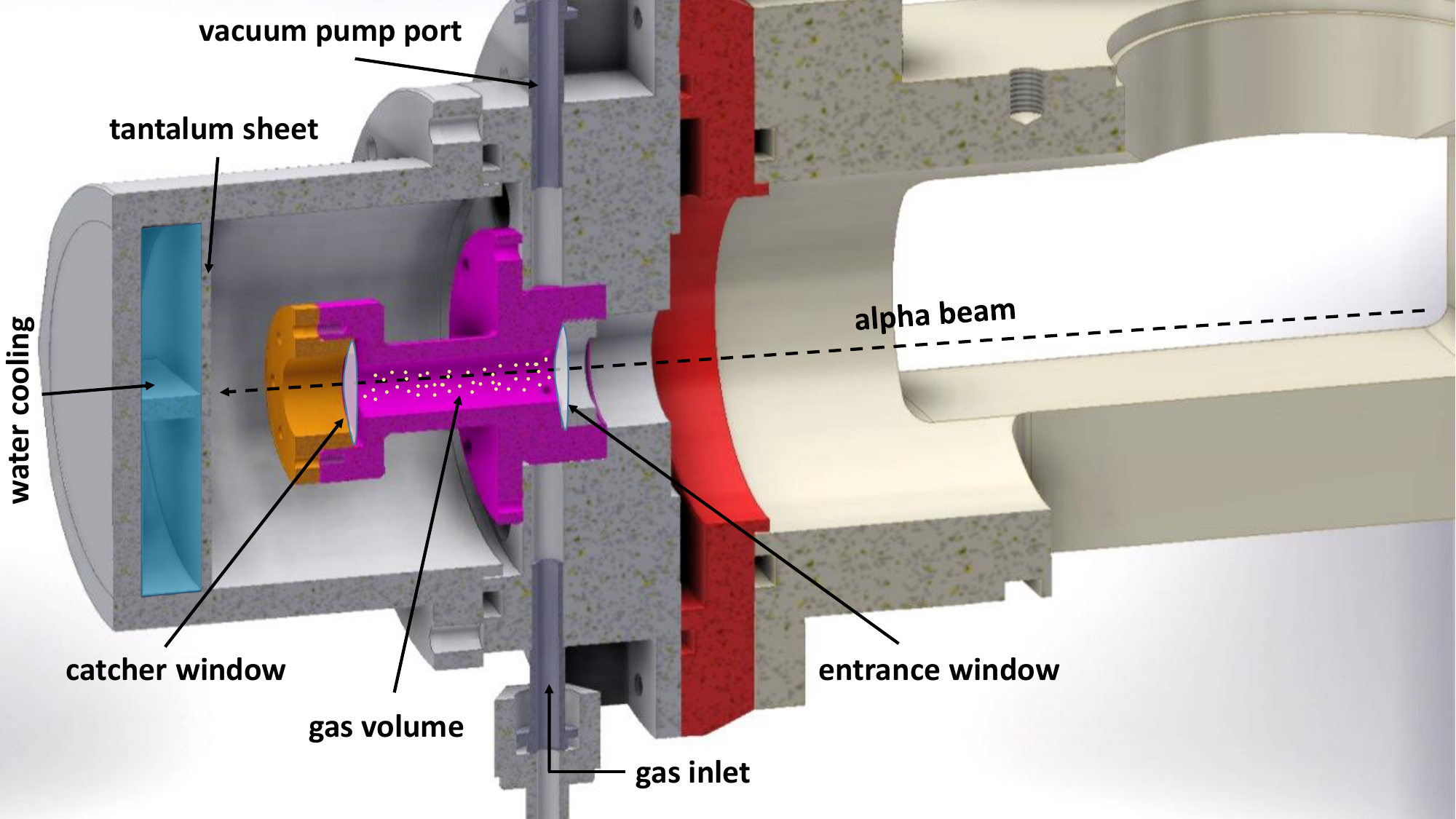}
\caption{\label{fig:setup} Cutaway technical drawing of the gas-cell used for the irradiations. The complete cell is surrounded by the beam-line vacuum. See text for details.}
\end{figure}

The irradiations were performed by the Atomki \mbox{MGC-20} cyclotron \cite{Biri21-EPJP}. The activation chamber containing the thin window gas-cell acted as a Faraday-cup.
A voltage of $-300$\,V was applied to an aperture at the entrance of the chamber to eliminate the effect of any secondary electrons that may be generated in the last beam defining aperture or from the target. Since the target gas was within the Faraday-cup, charge exchanges did not affect the current measurement. This allowed the determination of the number of bombarding particles via charge integration. A pressure gauge was also connected to the cell, and the pressure data were saved every 10 minutes. During the irradiation the pressure in the cell increased steadily (by about 15\% observed until the end of a given irradiation), and a slow decrease was observed after the irradiation was stopped (1-2\% within a few days). 
Since the whole cell was surrounded by vacuum, in case of a foil or o-ring failure only pressure drop shall be observed, thus the pressure increase is considered to be a temperature effect (few \% increase is consistent with the few degree temperate increase), and mainly gas desorption from the foils and cell walls. The pressure increase was always more significant when the cell was exposed to air for a longer time before irradiation. None of these effects alter the number of active target atoms. The energy loss of the beam inside the gas volume was in the order of a few tens of keV. Considering the total pressure increase caused by air desorption, this small extra gas amount alters the center-of-mass energy by less than 0.1\%, well within the initial beam energy uncertainty.
The unreacted beam was dumped in a water cooled tantalum cap. The drawing of the gas target chamber is shown in \fig{fig:setup}.

The beam intensity was monitored by a charge integrator combined with a multichannel scaler. The accumulated charge was recorded in every 60 seconds, which allowed the current variation be to taken into account in the data analysis.
The length of the irradiations was between 10 and 26 hours to create the adequate activity in the catcher foil. The electric beam current of the doubly charged $\alpha$ particles varied between $0.6-1.2$ $\mu$A depending on the actual performance of the accelerator.

\subsection{\label{sec:count} $\gamma$-ray detection}

During the irradiation, all the $^7$Be nuclei were implanted in the catcher foil. SRIM simulations \cite{srim} were carried out to investigate the possible backscattering or stopping of $^7$Be in the gas volume. Both of these effects were found to be negligible ($<$\,0.01\%). 
Due to the reaction kinematics, the $^7$Be was created in a cone with a maximum of 26.0\,mrad opening angel (in case of the highest energy irradiation). This would results in a maximum 2\,mm diameter spot on the catcher. Since the original $\alpha$ beam had a size of 5~mm, defined by the last aperture, this additional broadening of the $^7$Be distribution is not significant.
After the irradiations, the catcher foils were removed from the gas-cell, and were placed in front of a high-purity germanium (HPGe) detector with a sample detector distance of 1~cm. Typically, the $\gamma$-ray counting was started with a cooling time of at least one day after irradiation, because there was some significant beam-induced short lived activity found in the catcher immediately after irradiation.
Two HPGe  detectors were employed for $\gamma$-ray countings depending on their availability.
A Canberra GL2015R type Low Energy Germanium Detector (LEGe) \cite{LEGe} with standard dipstick cryostat and a Canberra GR10024 N-type detector with Ultra Low Background (ULB) \cite{ULB} cryostat. The detectors were surrounded by lead shielding including inner layer of Cd and Cu, with which their sensitivity become comparable at E$_\gamma$ = 477.6 keV \cite{Szucs14-AIPConf}.
In most of the cases a given sample was measured by both of the detectors, and the yields obtained were within statistical uncertainty. These data points obtained with the two detectors have most of their systematic uncertainties common, thus the evaluation was performed with the spectra measured with the ULB detector, since it had better statistics. Countings were performed in several cycles, a given sample was in the detector setup for a few days, then it was placed back after about a week of waiting time. In this way the $^7$Be decay was followed in each sample, which was found to be compatible with the expectations assuming the literature half-life of $^7$Be. The total counting time of a given sample was $4-22$ days, depending on the activity, to reach $2-2.5\%$ statistical uncertainty.

\begin{figure}[b]
\includegraphics[width=0.99\columnwidth]{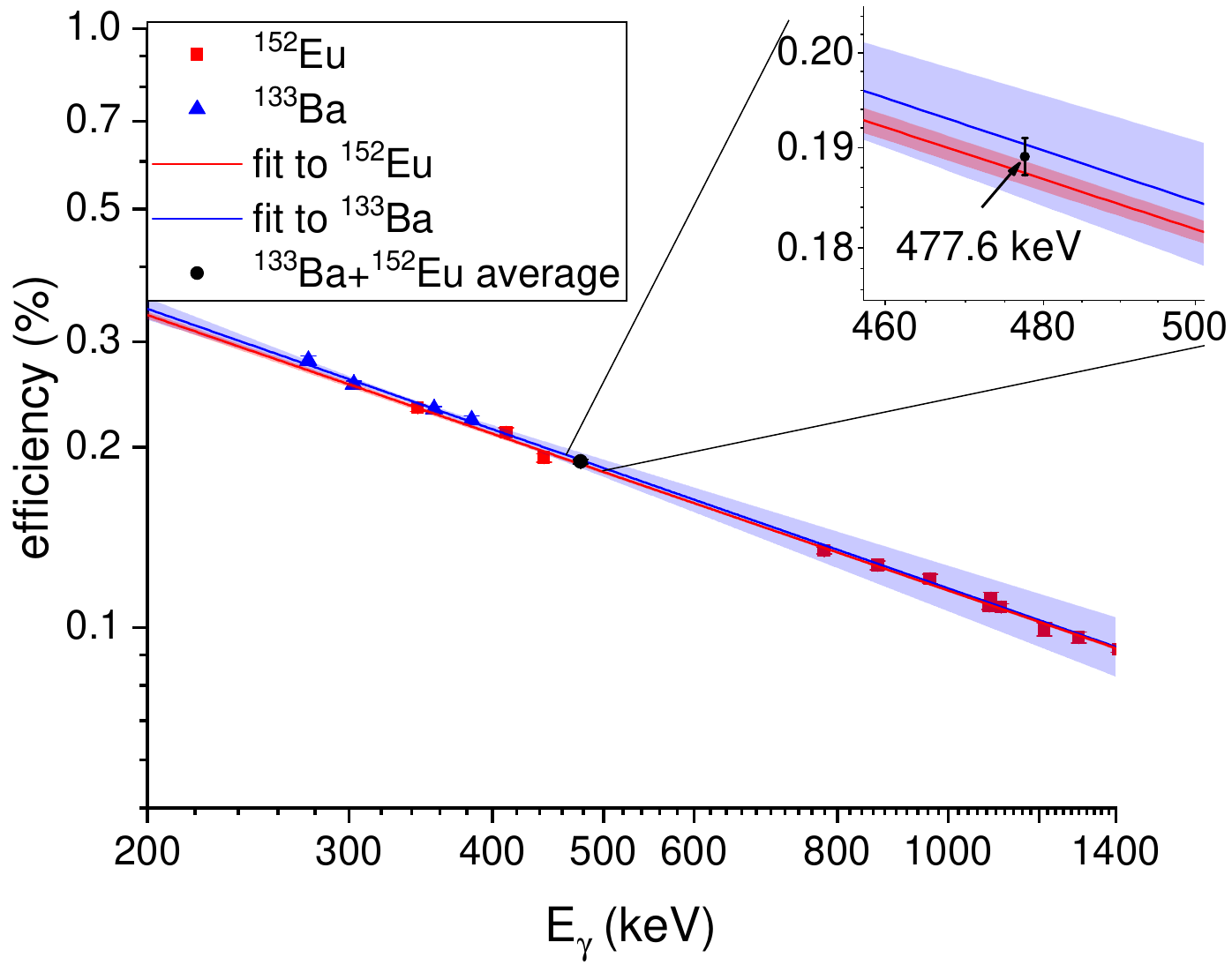}
\caption{\label{fig:eff} Efficiency determination of the ULB $\gamma$-ray detector at 27 cm source-detector distance. The extrapolation and interpolation to 477.6 keV is shown in the inset together with the averaged value finally used in the analysis. The shaded areas are the 1$\sigma$ confidence intervals of the fits.}
\end{figure}

The efficiency calibration of the detectors were performed with a custom-made $^7$Be single-line source produced via $^7$Li(p,n)$^7$Be reaction. The source was created in the same irradiation setup, thus the proton beam was collimated to a 5~mm spot, where the activity was created evenly in the target material. With this method a calibration source geometry was achieved which was similar to the extended activity distribution in the catcher.
The $^7$Be source activity was measured with high precision at the ULB detector using 27 cm source-detector distance. 
Commercial calibration sources of known activities ($^{152}$Eu and $^{133}$Ba) were used for the determination of detector efficiency-energy function at this distance. With these multi-line sources, direct close geometry calibration would be affected by the true coincidence summing, which was avoided in this way.
Using high intensity $\gamma$ transitions from both sources, the detection efficiency was determined, then a log-log linear function was fitted separately for the values obtained by both sources. In case of $^{133}$Ba, the detection efficiency at the E$_\gamma$ = 477.6 keV $^7$Be line was extrapolated, in case of $^{152}$Eu interpolation was possible (see \fig{fig:eff}).
The determined detection efficiency at the $^7$Be line was in mutual agreement. Taking into account the normalisation uncertainty stemming from the source activities, the weighted average value was used later in the analysis, carrying only 1\% uncertainty, which in turn gives the precision of the $^7$Be source activity.

The detection efficiencies at 1 cm source-detector distances for E$_\gamma$ = 477.6 keV of both detectors were then determined with the precisely calibrated $^7$Be single-line source with high accuracy (1.5\%). 

The detector efficiency was measured again after the $\gamma$-ray countings with the same calibration sources and with another freshly produced $^7$Be source to test the stability of the system. There were no significant change in the efficiency (in the order of 0.3\%, within the statistical uncertainty), thus the error-weighted average of the two efficiency results was used in the analysis.

\subsection{\label{sec:analysis} Data analysis}

\begin{figure}[b]
\includegraphics[width=0.99\columnwidth]{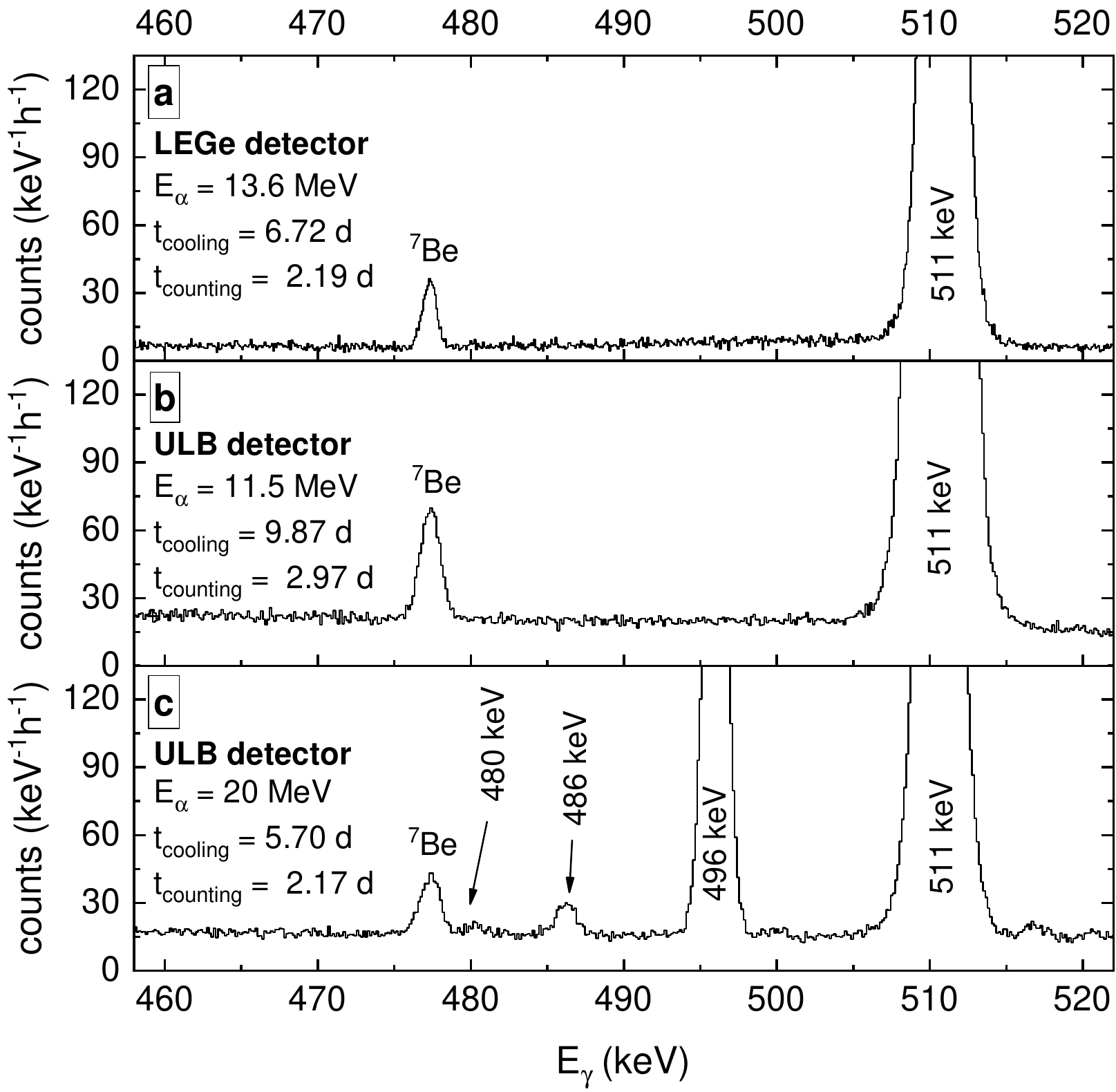} 
\caption{\label{fig:spe} On the top, in panel (a), a typical spectrum taken by the LEGe detector, in the middle panel (b) another one taken by the ULB on samples from lower energy irradiations, while on the bottom, in panel (c), a typical spectrum after a higher energy irradiation is shown. In all cases the peak from $^7$Be is clearly separable, even though in the high energy irradiation more parasitic activity was created in the exit foil. See text for details.}
\end{figure}

Typical $\gamma$-ray spectra are shown in \fig{fig:spe}: the spectra taken after lower energy irradiations show less beam induced background from parasitic reactions in both detectors (\fig{fig:spe} (a) and (b)). A spectrum taken after a higher energy irradiation, which caused more parasitic activity in the foil, thus featuring more contaminant peaks in the spectrum is also displayed (\fig{fig:spe} (c)).

Despite the high purity (99.99\% Al) of the foils, they contain some impurities on ppm level (Cu, Fe, Mg, Si...), due to the manufacturing process. These impurities in the foil may also undergo $\alpha$-particle induced reactions. The half-life of the resulting radioactive nuclei is usually less than 1 day. The most probable reactions are ($\alpha$,n), but at such high energies several reaction channels can be open, such as \rann reactions. The latter is of great importance, since the $^{54}$Fe nucleus present in the foil (among others) is the target of such a reaction, with a reaction threshold of 17.2~MeV. The $^{54}$Fe\rann reaction produces $^{56}$Ni, which is radioactive and has a half-life of almost 6 days. During the decay of $^{56}$Ni, $\gamma$ photons with energy of E$_\gamma$ = 480 keV are emitted. Due to the finite energy resolution of the detector, this manifests as a side peak/shoulder of the $^7$Be peak with its energy of E$_\gamma$ = 477.6~keV. This small structure was considered in the peak area determination for the $E_\alpha = 17.5-20$~MeV irradiations.
In addition, a prominent peak at E$_\gamma$ = 496~keV and a smaller one at E$_\gamma$ = 486~keV are visible in the spectra after the high energy irradiations (see \fig{fig:spe} bottom panel). Even though this directly does not affect our peak of interest, such parasitic peaks are not expected from reactions on the foil impurities. From their intensity ratio and half-life (and from other observed peaks) the source of these peaks was identified as $^{131}$Ba. This isotope was created via the  $^{129}$Xe\rann reaction, which has huge cross section ($0.3-0.5$ barn) above its threshold of 16.2 MeV \cite{TALYS-V19}. These parasitic peaks were visible only in the spectra taken after the irradiation at and above $E_\alpha =17.0$~MeV corresponding to 16.5 MeV effective beam energy behind the entrance foil. Trace xenon impurity on the ppm level was enough to create the observed amount of activity. Since the gas handling part of the gas-cell was previously used with natural Xe gas, despite the evacuation, a trace amount of xenon was trapped and mixed into the helium used for our experiments. 
In principle $^{131}$Ba can also be created via the $^{128}$Xe\ran reaction above the reaction threshold of 9~MeV, however this production is insignificant, because $^{128}$Xe has more than one order of magnitude lower natural abundance than $^{129}$Xe, and orders of magnitude lower $^{131}$Ba production cross section \cite{Rauscher00-ADNDT}.

The $^7$Be peak area was determined by fitting the spectrum with a lognormal function assuming linear background below the peak. The slight low energy wing of the peaks due to incomplete charge collection in the germanium crystal is taken into account in this way. The asymmetry is small, assuming Gaussian peak would change the peak area within the statistical uncertainty. The peak area was then corrected for detector dead time, and random coincidence loss effects, both on the 0.1\% level.

\begin{figure*}[p]
\includegraphics[width=0.75\paperwidth]{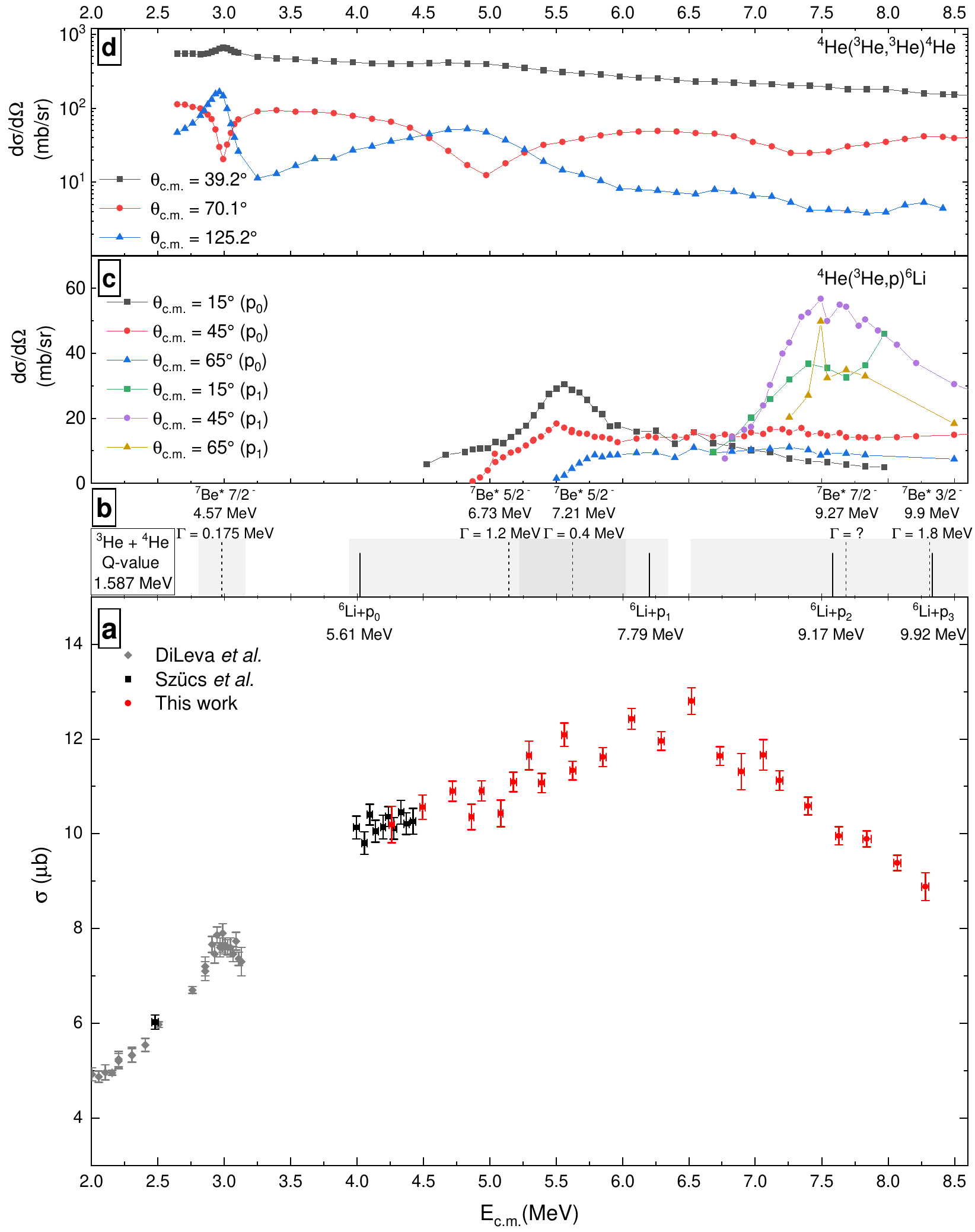}
\caption{\label{fig:XS} In the bottom panel (a) the present reaction cross sections together with data from previous works \cite{DiLeva09-PRL,Szucs19-PRC} are displayed. Only the statistical uncertainty of the data points is plotted. In panel (b) the $^7$Be levels are shown with dashed central lines, and shaded total widths taken from the most recent compilation \cite{Tilley02-NPA}. The proton separation thresholds are indicated with solid lines. Each of the energies are shifted for the plot with the reaction Q-value to match the experimental energy scale. On the top panels selected differential cross section data from the literature for the $^3$He($\alpha$,p)$^6$Li reaction \cite{Spiger67-PR} (c) and for the elastic scattering \cite{Spiger67-PR} (d) are plotted. The lines are just to guide the eye.}
\end{figure*}

The statistical uncertainty of $\gamma$ countings was generally $2-2.5\%$. The uncertainty of the $^3$He target thickness was between $2.5 - 2.7\%$. One of the dominant uncertainties was the cell length uncertainty of 1~mm amounting to 2.4\%. This is a conservative upper limit including the uncertainty of the length measurement (0.2~mm), and the bending of the foils in the order of 0.3~mm, when exposed to pressure difference \cite{Halasz16-PRC}. Further uncertainties were considered such as beam heating effect (between $0.6-1\%$) \cite{Marta06-NIMA}, the cooling water temperature, which defines initial gas temperature (0.7$\%$) and the pressure in the cell (0.3$\%$). The uncertainty of the bombarding particle flux was assumed to be 3$\%$. Taking the quadratic sum of the above partial uncertainties, the reaction cross section was determined with an accuracy of $4.6-5.8\%$.

\begin{table}[t]
\vspace{-2mm}
\caption{The obtained cross section dataset with the statistical and systematic uncertainties, and with the corresponding effective center-of-mass energies.}
\label{tab:res}
\center
\begin{ruledtabular}
\begin{tabular}{c c c}									
$E_{\alpha}$ 		&	$E_{c.m.}$\,$\pm$\,$\Delta$$E_{c.m.}$	&	$\sigma$\,$\pm$\,$\Delta$$\sigma_{stat}$\,$\pm$\,$\Delta$$\sigma_{syst}$ \\
(MeV)	& (MeV)	&	($\mu$b)	\\
\noalign{\smallskip}\colrule\noalign{\smallskip}		
11.0\footnotemark[1]						&	4.284\,$\pm$0.014		&	10.19\,$\pm$0.36\,$\pm$0.44 \\
11.5										&	4.517\,$\pm$0.015		&	10.56\,$\pm$0.20\,$\pm$0.46 \\
12.0										&	4.742\,$\pm$0.015		&	10.89\,$\pm$0.21\,$\pm$0.47 \\
12.0\footnotemark[2]						&	4.881\,$\pm$0.015		&	10.43\,$\pm$0.28\,$\pm$0.45 \\
12.5										&	4.962\,$\pm$0.016		&	10.90\,$\pm$0.21\,$\pm$0.47 \\
12.5\footnotemark[2]						&	5.104\,$\pm$0.016		&	10.35\,$\pm$0.27\,$\pm$0.45 \\
13.0										&	5.202\,$\pm$0.017		&	11.09\,$\pm$0.21\,$\pm$0.48 \\
13.0\footnotemark[1]$^,$\footnotemark[2]	&	5.315\,$\pm$0.016		&	11.65\,$\pm$0.30\,$\pm$0.51 \\
13.5										&	5.415\,$\pm$0.017		&	11.07\,$\pm$0.20\,$\pm$0.48 \\
13.6\footnotemark[1]$^,$\footnotemark[2]  	&	5.582\,$\pm$0.017		&	12.09\,$\pm$0.25\,$\pm$0.52 \\
14.0										&	5.648\,$\pm$0.018		&	11.33\,$\pm$0.20\,$\pm$0.49 \\
14.5										&	5.877\,$\pm$0.018		&	11.62\,$\pm$0.20\,$\pm$0.50 \\
15.0										&	6.093\,$\pm$0.019		&	12.42\,$\pm$0.22\,$\pm$0.54 \\
15.5										&	6.317\,$\pm$0.020		&	11.96\,$\pm$0.20\,$\pm$0.52 \\
16.0										&	6.544\,$\pm$0.020		&	12.79\,$\pm$0.27\,$\pm$0.56 \\
16.5										&	6.759\,$\pm$0.021		&	11.64\,$\pm$0.20\,$\pm$0.51 \\
16.6\footnotemark[2]						&	6.917\,$\pm$0.021		&	11.31\,$\pm$0.22\,$\pm$0.49 \\
17.0\footnotemark[1]$^,$\footnotemark[2]	&	7.084\,$\pm$0.021		&	11.66\,$\pm$0.28\,$\pm$0.51 \\
17.5										&	7.209\,$\pm$0.022		&	11.12\,$\pm$0.21\,$\pm$0.48 \\
18.0										&	7.423\,$\pm$0.023		&	10.58\,$\pm$0.19\,$\pm$0.46 \\
18.5										&	7.656\,$\pm$0.023		&	\phantom{1}9.95\,$\pm$0.19\,$\pm$0.43 \\
19.0										&	7.868\,$\pm$0.024		&	\phantom{1}9.89\,$\pm$0.17\,$\pm$0.43 \\
19.5										&	8.097\,$\pm$0.025		&	\phantom{1}9.38\,$\pm$0.17\,$\pm$0.41 \\
20.0										&	8.309\,$\pm$0.025		&	\phantom{1}8.88\,$\pm$0.22\,$\pm$0.39 \\
\end{tabular}	
\end{ruledtabular}
\footnotetext[1]{Measured by the LEGe detector only.}
\footnotetext[2]{Thinner entrance foil was used during the irradiation.}		
\end{table}
\vspace{-3mm}

The uncertainty in the center-of-mass energy was between $0.3-0.5\%$ which is the quadratic sum of the cyclotron energy uncertainty of 0.3\% and the uncertainty caused by the energy loss in the entrance foil ($0.1-0.4\%$). This latter stems from the uncertainty of the Al foil thickness and the stopping power uncertainty. The value of the stopping power used in the calculation is 0.985 times that of in the SRIM \cite{srim} tables. This is due to the fact, that in the 10.5 - 20 MeV $\alpha$-energy range, which is the range of the $\alpha$ particles used for the irradiations, there is only one experimental stopping power dataset \cite{Andersen77-PRA}. The SRIM curve describes well the energy dependence of the high accuracy (0.6\%) data, but the absolute magnitude of the SRIM curve is 1.5\% higher. This scale shift was applied in our calculations, and a conservative uncertainty of 1.5\% was assumed.
The energy loss in the target gas was $25-44$ keV, assuming 4.4\% uncertainty of the stopping power according to SRIM \cite{srim}, this amounted to a negligible (max. 0.02\%) uncertainty of the effective reaction energy. Since the cross section is roughly constant within the above mentioned target thickness, the effective reaction energy was taken as the energy at the middle of the target.

The experimental cross section results together with the effective center-of-mass reaction energies are summarized in \tab{tab:res} and displayed in \fig{fig:XS}.

\section{\label{sec:results} Discussion}
The obtained excitation function is shown in \fig{fig:XS}. The gas-cell in the present work is different from the one used in Ref.~\cite{Szucs19-PRC}, thus an overlapping point was taken at about $E_{c.m.} = 4.3$~MeV as a cross validation. The new data point is in perfect agreement with the previous one.

In \fig{fig:XS} differential elastic scattering and $^3$He($\alpha$,p)$^6$Li reaction cross sections from Ref.~\cite{Spiger67-PR} are also plotted for selected angles. A complete compilation of the available datasets are beyond the scope of this work, here only the major features are highlighted.

In the present dataset no structures are visible around the known $^7$Be levels, while the 6.73 MeV level appears in the elastic scattering data, and the 7.21 MeV level forms a structure in the $^3$He($\alpha$,p$_0$)$^6$Li dataset. This suggest a marginal $\gamma$ widths for these levels beside sizable particle widths. Similarly, the two other levels in the investigated energy range show structures in the particle channels, but not visible in the radiative capture dataset. Above its threshold, the $^3$He($\alpha$,p$_1$)$^6$Li cross section becomes dominant, this is the energy range, where the present cross section starts to drop. The $^3$He($\alpha$,p$_1$)$^6$Li cross section peaks at the $^3$He($\alpha$,p$_2$)$^6$Li reaction threshold, from which point supposedly the latter reaction becomes dominant, however no experimental data are available for that reaction channel. Additionally, the $^3$He($\alpha$,p$_2$)$^6$Li reaction threshold energy is close to a broad 7/2$^-$ level in $^7$Be, thus that may also cause the structure of the $^3$He($\alpha$,p$_1$)$^6$Li cross section.

\begin{figure}[b]
\includegraphics[width=0.99\columnwidth]{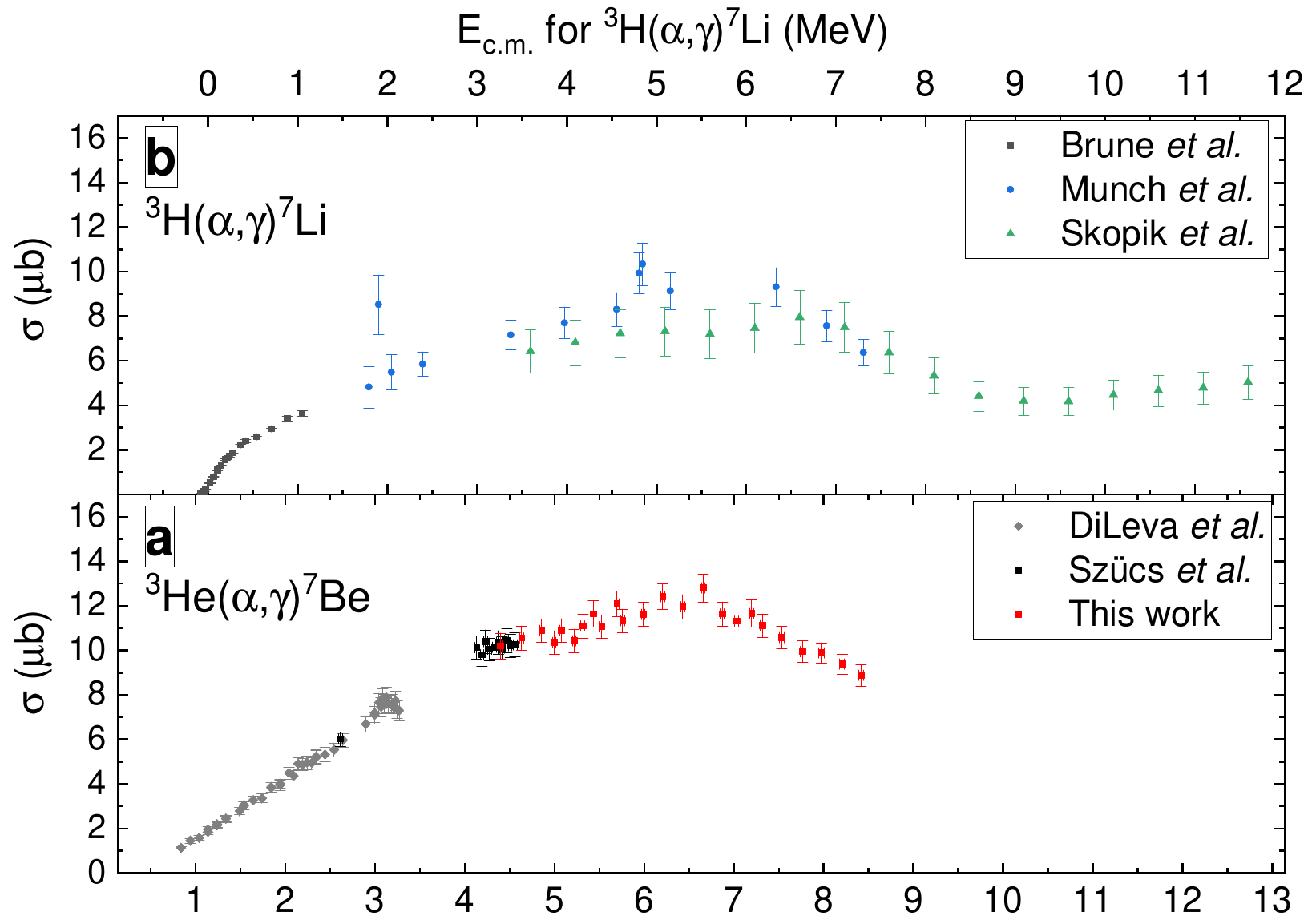}
\caption{\label{fig:7Li} The present data (bottom panel (a)) compared to the $^3$H($\alpha$,$\gamma$)$^7$Li$_{GS}$ reaction cross section (top panel (b)) from \cite{Brune94-PRC, Skopik79-PRC, Munch20-PRC}. The x-axes are aligned to account for the \mbox{Q-value} difference of the reactions.}
\end{figure}

Comparing the present data to the cross section of the mirror reaction i.e. $^3$H($\alpha$,$\gamma$)$^7$Li$_{GS}$, shows remarkable common features (see \fig{fig:7Li}). The higher energy $^3$H($\alpha$,$\gamma$)$^7$Li$_{GS}$ data were obtained by $\gamma$ induced breakup reaction on $^7$Li \cite{Skopik79-PRC, Munch20-PRC}. The cross section of the measured $^7$Li($\gamma$,$\alpha$)$^3$H reaction is converted to the plotted $^3$H($\alpha$,$\gamma$)$^7$Li$_{GS}$ using the principle of detailed balance.
A similar broad structure between 4 and 9 MeV is visible in both reactions. Because of the maximum energy of our accelerator, the new dataset does not cover higher energies, where the \cite{Skopik79-PRC} data become constant. Investigations up to $E_{c.m.}= 13$ MeV is recommended to confirm this similar behavior towards higher energies.

\begin{figure}[b]
\vspace{-2mm}
\includegraphics[width=0.99\columnwidth]{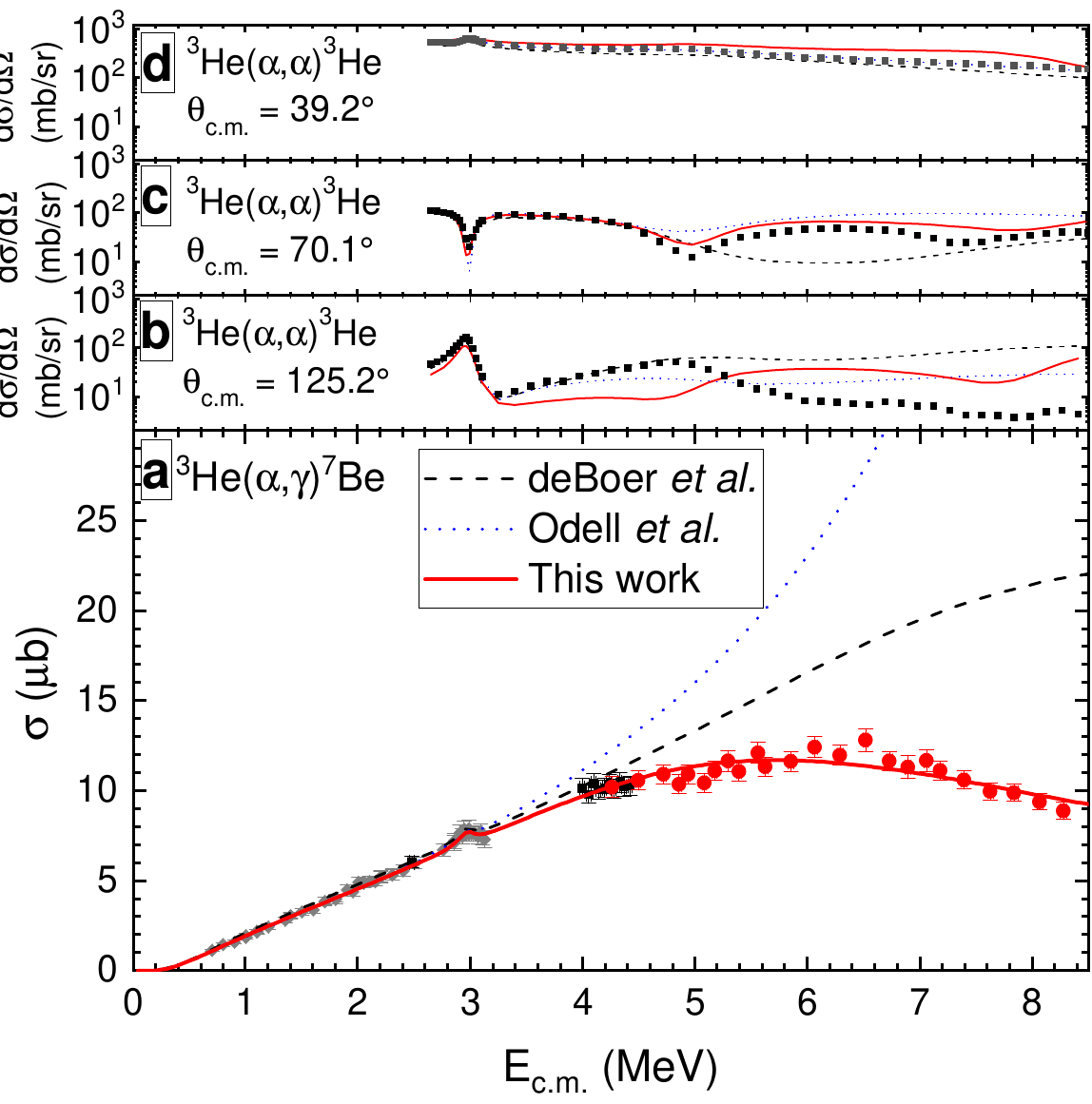}
\caption{\label{fig:R-matrix} On panel (a) the present results and literature datasets from Refs. \cite{DiLeva09-PRL,Szucs19-PRC} together with R-matrix calculations from previous works \cite{deBoer14-PRC, Odell22-FP} and with the adjusted parameters from this work are displayed. Panels (b)-(d) show elastic scattering excitation functions from Ref. \cite{Spiger67-PR} at selected angles together with the R-matrix fits.}
\end{figure}

Finally, the present data are also compared to previous literature R-matrix fits \cite{deBoer14-PRC, Odell22-FP} using the AZURE2 code \cite{Azuma10-PRC} (see \fig{fig:R-matrix}). Those fits considered only data below $E_{c.m.}=3$ MeV, and used many background poles well outside the range of the data. In \tab{tab:r-matrix} the spin, parity, energy, and $\alpha$ widths of the levels considered in the previous R-matrix fits and in the present one are shown. 
Ref.~\cite{deBoer14-PRC} used altogether 7 poles all placed at $E_{x}=11$ MeV, but the $5/2^-$ one, which was placed at $E_{x}=7$ MeV to account for known levels around this energy. Ref.~\cite{Odell22-FP} used altogether 6 poles skipping the $7/2^-$, and placed positive and negative parity poles at different energies.
As can be seen in \fig{fig:R-matrix}, the two fits start to deviate from each other already from $E_{c.m.}=3$ MeV, and completely miss the new data sets, as they were not intended to be used in that energy range.

\begin{table}[b]
\vspace{-4mm}
\caption{Levels used in the R-matrix analysis of this and previous works. Level energies and their $\alpha$ widths are in MeV.  In case of the ground ($E_{x} = 0$) and first excited states ($E_{x} = 0.429$), ANC-s in fm$^{-1/2}$ are displayed instead of level widths. For the present work, fitted widths are displayed in {\bf bold}, the other values were kept fixed for the fitting. The values marked in {\it italic} numbers are taken from the most recent compilation \cite{Tilley02-NPA}. }
\label{tab:r-matrix}
\center
\begin{ruledtabular}
\begin{tabular}{c c c c c c c}									
$J^{\pi}$ 	&	$E_{x}$ & $\Gamma_{\alpha}$ & $E_{x}$ & $\Gamma_{\alpha}$ & $E_{x}$ & $\Gamma_{\alpha}$ \\
		&	\multicolumn{2}{c}{\cite{deBoer14-PRC}} &\multicolumn{2}{c}{\cite{Odell22-FP}} & \multicolumn{2}{c}{This work} \\
\noalign{\smallskip}\colrule\noalign{\smallskip}		
1/2$^-$	 &  0.429	&  3.6  &  0.429	& \it3.0	  & \it 0.429 & 3.0	\\
1/2$^-$  &			&	 	  &			& 		  & \it 10	 & \bf 2.485	\\
1/2$^-$  & 11		& 12.1	  & 21.6		& 62.53	  & \it 17	 & \it 6.5 \\
1/2$^+$ & 11		& 19.3	  & 14		& 20.0	  & \bf 7.53	 & \bf 7.86	\\
3/2$^-$  &  0		&  3.7	  &  0		&  3.98 & \it 0	 &  3.98	\\
3/2$^-$  & 11		& 12.0	  & 21.6		& 19.64	  & \it 9.9	 & \it 1.8 \\
3/2$^+$ & 11		& 11.1	  & 12		& 7.15	  & 15	 & \bf 1.51 \\
5/2$^-$  & 7		& 2.9	  & 7		& 2.7	  & \it 6.7	 & \it 1.2 \\
5/2$^+$ & 11		& 11.1	  & 12		& 3.77	  & 15	 & \bf 3.26 \\
7/2$^-$  &  4.56		&  0.15	  &  4.59		&  0.163 & \it 4.57	 & \it 0.175 \\
7/2$^-$ & 11		& 8.8	  & 			& 		  & \it 9.27	 & \bf 0.000 \\
\end{tabular}	
\end{ruledtabular}					
\end{table}

A new R-matrix fit with limited number of reaction channels (radiative capture and elastic scattering involving only $^3$He and $^4$He), and datasets was performed. Hereafter we discuss this new limited fit.
A comprehensive R-matrix fit including multiple reaction channels such as $^3$He($\alpha$,p$_0$)$^6$Li, $^3$He($\alpha$,p$_1$)$^6$Li, $^6$Li(p,$\gamma$)$^7$Be, $^6$Li(p,p)$^6$Li and more datasets is beyond the scope of this paper.

For the radiative capture channel, the new data and two previous datasets \cite{DiLeva09-PRL,Szucs19-PRC} were considered. For the scattering channel one dataset from Ref.~\cite{Spiger67-PR} was used to better constrain the $\alpha$ widths.
As a starting point, the levels and $\alpha$ widths from the most recent compilation \cite{Tilley02-NPA} were used and the $\gamma$ widths and ANC-s were taken from Ref.~\cite{Odell22-FP}. Since no low energy data was used, the ANC-s were kept fixed.
The energy of the levels was also fixed, the only exception was the $1/2^+$ level which was initially placed to $E_{x}=7.5$ MeV and fitted in the seek of describing the apparent broad structure. The $\alpha$ widths of the positive parity poles were varied, so were the $\gamma$ widths of them. The partial $\gamma$ width to the ground and to the first excited state was not constrained here, because no partial cross sections were used for the fit. The  $1/2^-$ level at $E_{x}=10$ MeV is indicated in the compilation only as a possible level marking it as "broad".
The resulting fit is plotted in \fig{fig:R-matrix}. The energy of the $1/2^+$ level did not changed significantly from its initial value, while the fit attributed an $\alpha$ width of ~8 MeV to this level.

The fit describes the trend of the data quite well, however the scattering data especially in backward angles are poorly reproduced. Nevertheless, the positive parity state in the range of the data did not make significant change in the low energy behavior of the R-matrix fit, their trend staid the same. Since no partial cross section data were used in the present fit, the ground state and first excited state ANC-s cannot be constrained separately. Thus, they were fixed to the values of previous works, resulting no change in the extrapolated zero energy cross section value.

\section{\label{sec:sum} Summary}

The cross section of the $^3$He($\alpha,\gamma$)$^7$Be reaction was measured for the first time over the energy range of $E_{c.m.}=4.3-8.3$ MeV, with 0.2 MeV energy step, using the activation technique.
The known $^7$Be levels cause no prominent features in the excitation function. However, the overall shape of the obtained cross section indicates a broad structure peaking at $E_x=7.5$ MeV $^7$Be excitation energy. A similar structure is visible in the $^3$H($\alpha$,$\gamma$)$^7$Li mirror reaction. 

A limited R-matrix fit was performed using only few additional radiative capture datasets \cite{DiLeva09-PRL,Szucs19-PRC} and one elastic scattering dataset \cite{Spiger67-PR}. The energies and widths of most of the levels were kept fixed, only the parameters of the positive parity poles, required for the description of the low energy behavior of the cross section \cite{Kontos13-PRC,deBoer14-PRC, Odell22-FP} were varied. Treating the broad structure as a $1/2^+$ positive parity state, the fit nicely describes the capture data in the energy range of the new dataset. The new fit does not differ significantly at lower energies from fits in previous investigations. The description of the elastic scattering dataset is poor, however not much worse than in other works. It has to be mentioned here, that those previous works did not use data in the energy range of this study, and were not intended for extrapolation to this higher energy range.
A comprehensive R-matrix fit is recommended which would use other reaction channels and partial cross section datasets to better describe the level scheme of the $^7$Be nucleus. 
Similarly, the study of the cross section of the radiative capture at even higher energy is required to compare the energy dependence of the cross section with the upturn observed in the $^3$H($\alpha$,$\gamma$)$^7$Li mirror reaction.

\begin{acknowledgments}
The authors thank R. J. deBoer (University of Notre Dame) for valuable discussions. We thank the operating crews of the cyclotron accelerator for their assistance during the irradiations. This work was supported by \mbox{NKFIH} (OTKA FK134845 and K134197), New National Excellence Programs of the Ministry of Human Capacities of Hungary under nos. \mbox{\'UNKP-22-3-II-DE-31} and \mbox{\'UNKP-22-5-DE-428}, and by the European Union (ChETEC-INFRA, project no. 101008324). T.S. acknowledges support from the J\'anos Bolyai research fellowship of the Hungarian Academy of Sciences.
\end{acknowledgments}

\end{document}